\documentclass[a4paper,english,english,english,english,english,preprint,preprintnumbers,amsmath,amssymb]{revtex4}
\usepackage[T1]{fontenc}
\usepackage[latin1]{inputenc}
\usepackage{babel}
\usepackage{array}
\usepackage{graphicx}
\usepackage{color}

\makeatletter

\begin{document}

\title{Non-equilibrium dynamics of an active colloidal ``chucker''}

\author{C. Valeriani}\thanks{Author to whom correspondence should be addressed. Electronic addresses:
cvaleria@ph.ed.ac.uk}
\affiliation{SUPA, School of Physics and Astronomy, University of 
Edinburgh, Mayfield Road, Edinburgh, EH9 3JZ, Scotland}
%\thanks{Author to whom correspondence should be addressed. Electronic addresses:
%cvaleria@ph.ed.ac.uk}
%\email{cvaleria@ph.ed.ac.uk}

\author{R. J. Allen}\thanks{RJA and DM made an equal contribution to this work}
\affiliation{SUPA, School of Physics and Astronomy, University of 
Edinburgh, Mayfield Road, Edinburgh, EH9 3JZ, Scotland}
%\email{rallen2@ph.ed.ac.uk}
%\altaffiliation{equal contribution}

\author{D. Marenduzzo}\thanks{RJA and DM made an equal contribution to this work}
\affiliation{SUPA, School of Physics and Astronomy, University of 
Edinburgh, Mayfield Road, Edinburgh, EH9 3JZ, Scotland}
%\altaffiliation{equal contribution}

\begin{abstract}
We report Monte Carlo simulations of the dynamics of a ``chucker'': a colloidal particle 
which emits smaller solute particles from its surface, isotropically and at a constant rate $k_{\rm c}$. 
We find that the diffusion constant of the chucker increases for small $k_{\rm c}$, as recently predicted theoretically. 
At large $k_{\rm c}$ the chucker diffuses more slowly due to crowding effects. 
We compare our simulation results to those of a ``point particle'' Langevin dynamics scheme 
in which the solute concentration field is calculated analytically, and in which hydrodynamic effects 
arising from colloid-solvent surface interactions can be accounted for in a coarse-grained way. 
By simulating the dragging 
of a chucker, we obtain an estimate of its apparent mobility coefficient which violates the
fluctuation-dissipation theorem.  We also characterise the probability density profile for a 
chucker which sediments onto a surface which either repels or absorbs the solute particles, 
and find that the steady state distributions are very different in the two cases. 
Our simulations are inspired by the biological example of exopolysaccharide-producing bacteria, 
as well as by recent experimental, simulation and theoretical work on phoretic colloidal ``swimmers''.
\end{abstract}

\pacs{05.70.Fh,87.18.Hf}

\maketitle

\section{Introduction}\label{sec:intro}

Understanding the properties of colloidal systems which are intrinsically out-of-equilibrium as they are active, is an important emerging goal in soft matter physics, 
driven both by the potential for modelling motile cells as active colloidal ``swimmers'', 
and for designing self-propelled particles with novel nanotechnological applications~\cite{hydro1,Ramaswamy1,Ramaswamy2,hydro2,hydro3,AP,Lowen09}. 
Several mechanisms for colloidal self-propulsion have been proposed: these can be broadly grouped into those where 
the colloid exerts force on the surrounding fluid (as for swimming bacteria or other motile cells)~\cite{hydro1,Ramaswamy1,Ramaswamy2,hydro2,hydro3}, 
and those in which the colloid changes the chemical properties of the surrounding medium in an asymmetric way by catalysing a 
chemical reaction on its surface~\cite{osmoGol07,Sen1,Sen2} or by secreting some product~\cite{Golestanian09,Gol05,Brady08,Kapral,popescu}.

 In this paper, we consider a ``chucker'': a spherical colloidal particle which produces smaller ``solute'' 
particles at its surface. Our model colloid is simpler than most cases considered thus far~\cite{Gol05,Brady08,Kapral,popescu},
 since it ``chucks'' solute particles isotropically, and therefore does not display directional motion. However, the chucking
 drives our system  out of equilibrium, giving it interesting dynamical properties. We find that the effective diffusion constant 
of the chucker increases with the chucking rate $k_{\rm c}$ for low $k_{\rm c}$ ,  and eventually decreases for high $k_{\rm c}$ due to self-crowding. 
We further simulate the dragging of a chucker, to obtain a measurement of its effective mobility. Comparing this to the effective
 diffusion constant allows us to show that the fluctuation-dissipation theorem does not hold in this non-equilibrium
system, and that the deviations from the FDT are non-monotonic in the chucking rate.  Finally, we consider the steady-state positional probability 
distribution function for a chucker which sediments against a planar surface, in the case where the surface is ``hard'', and in the case where the surface absorbs the solute particles.

 Although our model is designed to be generic, our work is inspired by the observation that many bacterial cells secrete 
extracellular polysaccharides \cite{sutherland1982,sutherland1994}, in some cases in large quantities (for example the
 bacterium {\em{Xanthomonas campestris}} can produce xanthan polymer at a rate  as high as $10^4$ polymer molecules per cell per 
second~\cite{xanthomonas1,xanthomonas2}). These exopolysaccharides can have important effects on the collective properties 
(e.g. phase behaviour) of assemblies of bacteria \cite{wilson_prep}. Our simulations suggest that in some cases 
exopolysaccharide production might also affect the diffusive motion of single bacterial cells. We note that the case 
of a chucker sedimenting  onto an absorbing surface might be relevant to exopolysaccharide-producing bacteria located close to a biofilm.

Several recent works have considered the dynamics of colloids which produce surface particles from a theoretical 
point of view~\cite{Gol05,Brady08,Kapral,popescu,Prost08} and using simulations \cite{Brady08}. The key concept 
here is that motion of the colloid can be generated by a concentration gradient of solute, produced by the 
colloid itself. Most previous work has considered the case where the colloid produces solute anisotropically across 
its surface, resulting in directional motion. In contrast, we consider the case where solute is produced isotropically,
 so that the solute concentration gradient is on average symmetrical about the colloid. However, as we show here, solute 
production nevertheless has a strong effect on the dynamical fluctuations of the colloid. The case of an isotropic chucker 
has recently been considered theoretically by Golestanian \cite{Golestanian09}. Using scaling theory and linearised hydrodynamics, he predicts inertial 
and diffusive dynamical regimes, separated by an anomalous regime for intermediate timescales in which the mean square 
displacement scales as the 3/2 power of time. He further predicts that in the diffusive regime, the effective diffusion 
constant scales linearly with the chucking rate. In this paper, we test these predictions using numerical simulations.

The question of how a solute concentration gradient influences colloid motion has some subtleties. From one point of view, 
both the colloid and the solute particles are large compared to the solvent particles; Brownian Dynamics (BD) simulations, 
in which the solvent is replaced by a stochastic force on the particles, are generally believed to work well in this regime \cite{bd3,bd6,bd7,bd8}. 
In such a BD representation, a concentration difference of solute across the colloid surface would produce a force on the 
colloid ${\bf{f}}=-k_BT \int {\bf{\hat{n}}} c dS$ where $c$ is the local concentration of solute, the integral is over the 
colloid surface, ${\bf{\hat{n}}}$ is the normal to the surface, $k_B$ is the Boltzmann constant and $T$ is the temperature 
\cite{Brady08}. However, such a BD representation does not take account of the details of the solvent motion close to the chucker 
surface. This may be important, since, as emphasised recently by  J{\"{u}}licher and Prost \cite{Prost08}, hydrodynamic theory using 
linearised non-equilibrium thermodynamics predicts that relative motion between the colloid and solvent is dependent on 
the existence of a surface slip velocity. In a simple view, the density of the solvent is expected rapidly to compensate any imbalances 
in the solute concentration field to maintain a constant overall isotropic pressure, except in a very narrow region around the colloid
 where  molecular interactions between the colloid surface and the solute can generate a surface slip velocity. Equivalently, one might 
state that in order to generate a force, the configuration of solute particles relative to the colloid surface must change when the 
colloid moves. Because the solute moves with the fluid, such a configurational change can only be achieved if there is a surface slip velocity.

As discussed by Golestanian {\em{et al}}~\cite{Golestanian09,Gol05}, 
the colloid velocity can be expressed as the integral over the colloid surface of a slip velocity ${\bf{v_s}}=-\mu \nabla c$ 
where $c$ is the solute concentration, $\mu$ the surface mobility $\mu=k_BT \lambda^2/\eta$, $\eta$ being the viscosity and $\lambda$ a length-scale. In the non-permeable case, the latter 
is the Derjaguin length, describing the range of interactions between the colloid and solute. The length-scale $\lambda$ is then
 small, so that the linearised theory predicts that chucking will have little effect on colloidal motion. However, Adjari and 
Bocquet \cite{Ajdari06} have shown recently that for solutes smaller than the 
intrinsic slip length, a small degree of surface slip can lead to large enhancements of $\lambda$. Moreover, for colloids that 
are partially or fully permeable to solvent (osmiophoresis), anything between this result and a much larger colloid velocity, 
equal to the ``naive'' kinetic force  ${\bf{f}}=-k_BT \int {\bf{\hat{n}}} c dS$ multiplied by a mobility factor, is possible~\cite{Gol05,sucrose}. 
The motion of a semi-permeable rigid vesicle in response to a solute gradient has been analysed in a linearised framework by Anderson \cite{And2}.
In a later work, the applicability 
of the linearised theory to this type of problem has been called into question by an experimental study by Nardi {\em{et al}} \cite{sucrose}. 
These authors measured the motion of lipid vesicles in a solute concentration gradient, and found a drift velocity more than 3 
orders of magnitude faster than that predicted by the linearised theory, suggesting that a strong nonlinear coupling exists between 
osmosis and hydrodynamic flow.

Our simulations do not include explicit solvent particles. Instead, we model the chucker as a diffusing hard sphere and 
the solute particles as diffusing hard spheres which interact only with the chucker. We use  a dynamical Monte Carlo (MC) 
simulation scheme which has been shown to be equivalent to Brownian Dynamics for small trial displacements \cite{Cichocki90,bd3,bd6,bd7,bd8}. 
 By simulating a passive colloid in a fixed gradient of solute particles, we show that  the naive kinetic result for the colloid 
velocity, expected for Brownian Dynamics simulations, is recovered in our MC simulations, at least for small solute particles.
 To estimate the likely effects of neglecting the detailed solvent-chucker interactions, we complement our MC simulations with a Langevin 
dynamics approach,  in which 
the chucker is modelled by a diffusing point particle which experiences a drift force proportional to 
the local solute concentration gradient. The prefactor linking the force to the density gradient (which is proportional to $\lambda$) can be 
varied, mimicking the effects that would be expected with explicit solvent. The Langevin simulations are in qualitative 
agreement with our MC results; this agreement is also quantitative, for suitably chosen $\lambda$, for low chucking rates and when the solute particles are small compared to the colloid.

This work is structured as follows. In the next Section, we describe in detail our simulation model and methods. In Section III we report 
results for a diffusing chucker, a dragged chucker and a chucker close to a surface. Finally, Section IV contains a discussion and our conclusions.

\section{Model and Methods}\label{sec:method}

We model the chucker as a hard sphere  of radius $R_1$, with long-time diffusion coefficient $D_1$. 
With a rate $k_c$, the sphere produces (``chucks'') smaller particles (solute) isotropically at its surface. The 
solute particles have radius $R_2$ and long-time diffusion coefficient $D_2$, which is related to $D_1$ by Stokes' law: 
$D_2/D_1=R_1/R_2$.

In our simulations, the length unit is the chucker radius $R_1$ 
and we use two different values for the solute radius $R_2$: $R_2/R_1=0.1$ ($R_1/R_2=10$) or  $R_2/R_1=0.01$ ($R_1/R_2=100$). 

\subsection{Monte Carlo simulations}\label{sec:MC}

Our main simulation results are obtained using a dynamical Monte Carlo algorithm. In this algorithm, a particle 
is chosen at random and subjected to a trial displacement $\delta$, in a randomly chosen direction and with magnitude uniformly 
distributed in the range $0 \to 0.001R_1$ (for the chucker) and $0 \to 0.001R_1\sqrt{{R_1}/{R_2}}$ for the solute particles 
(this ratio of displacements is to ensure the correct ratio between the diffusion coefficients $D_1$ and $D_2$, as given by Stokes' Law). 
If this trial displacement results in an overlap between chucker and solute, it is rejected; otherwise it is accepted. One MC cycle 
consists of $N$ trial displacements, where $N$ is the current number of particles in the system. It has been shown in previous 
literature \cite{Cichocki90,And1,And2,bd3,bd6,bd7,bd8} that this algorithm is equivalent to Brownian Dynamics for small trial displacements. 
An MC cycle can be interpreted as the time unit in a dynamical trajectory, where the 
diffusion constant in reduced length units is given by $D_1/R_1^2=\frac{(\delta R_1/R_1)^2}{6}$ [MC cycles]$^{-1}$. 
At the beginning of every  MC cycle, a new solute particle is inserted with uniform probability of $k_c$, randomly at the surface of the chucker. Note that 
because the solutes do not interact, all particle insertions are accepted. 
All values for the chucking rate $k_c$ in our MC simulations are in units of [MC cycles]$^{-1}$. This dynamical MC scheme allows us 
to represent the solute particles as penetrable hard spheres and easily to insert new solute particles, both of which would be difficult in a standard Brownian Dynamics scheme.

The chucker and the solute particles interact according to the Asakura-Oosawa-Vrij (or penetrable hard sphere) 
model~\cite{AO1,AO2,vrij}: the solute particles cannot overlap with the chucker but do not interact with each other:
\begin{eqnarray}
v_{ss} &=& 0\\\nonumber
v_{sc} &=& 0 \qquad {\rm{if}}\,\, r_{sc} > R_1 + R_2\\\nonumber
v_{sc} &=& \infty \qquad {\rm{if}}\,\, r_{sc} \le R_1 + R_2\\\nonumber
%\qquad {\rm{otherwise}}
\end{eqnarray}
where $v_{ss}$ and $v_{sc}$ are the solute-solute and solute-chucker interaction energies, and $r_{sc}$ is the distance between 
the centres of the solute and chucker particles.

\begin{figure}[h!]
\begin{center}
\includegraphics[width=9.cm]{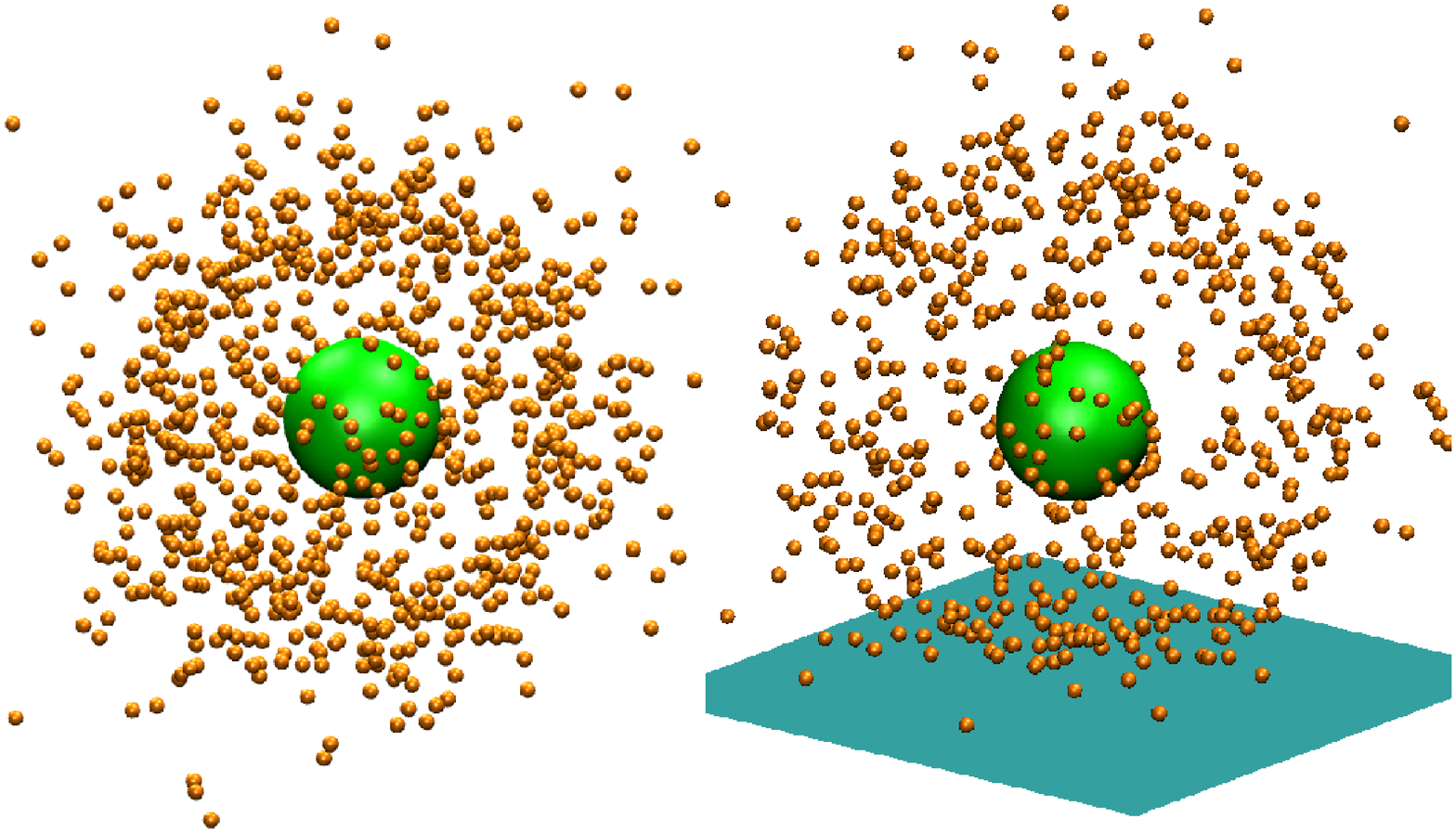}
\caption{Snapshot of a freely diffusing chucker (left), and a chucker close to a hard-planar surface (right).\label{fig:snapshots}}
\end{center}
\end{figure}

\begin{figure}[h!]
\begin{center}
\includegraphics[width=9cm,clip]{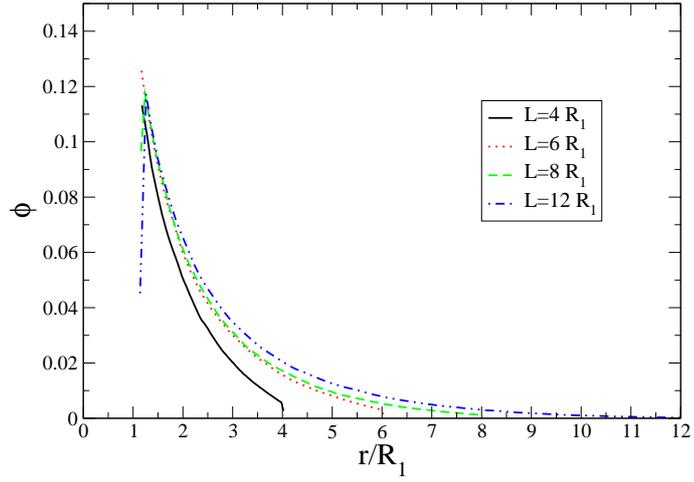}
\caption{Volume fraction ($\phi \equiv (4/3)\pi R_2^3 c$ (where $c$ is concentration))of solute particles surrounding a 
motile-chucker ($R_1/R_2=10$) with $k_{\rm c}R_1^2/D_2=600$
($k_{\rm c}=0.001[MC cycles]^{-1}$) and various box size $L$.\label{fig:boundaries}}
\end{center}
\end{figure}

Figure~\ref{fig:snapshots} shows snapshots from our MC simulations. To prevent the number of solute particles from growing 
indefinitely during the simulations, solutes need to be removed at some large distance from the chucker. Our system reaches a steady 
state when the rate of solute particle generation at the chucker surface equals the 
rate of solute absorption at the boundary. We achieve this in two different ways. For simulations with a fixed wall (as shown in the 
right-hand panel of Figure~\ref{fig:snapshots}), we define a fixed cubic simulation box and remove any solute particles which stray 
outside this box. For simulations in the bulk (as shown in the left panel of Figure~\ref{fig:snapshots}), we remove any solute 
particles that stray outside a cubic box with sides of length $2 L$, centred on the current position of the chucker. The box size is 
large enough to have no effect on the results reported here ($L=6R_1$). We have also verified that for our parameter set, the results are independent of the 
box size. Figure \ref{fig:boundaries} shows the steady-state concentration profile of solute particles 
around the (moving) chucker, for several different values of $L$: for $L \ge 6R_1$, the solute concentration profile is virtually independent 
of $L$.

\subsection{Langevin Dynamics simulations}\label{sec:lang}

To facilitate the simulation of larger systems, and to better understand the underlying physics, 
we have also simulated our system using an overdamped Langevin dynamics algorithm \cite{AT}, in which the 
chucker is represented as a point particle and the solute particles are represented as a space and time-dependent 
density field, the gradient of which provides a force on the chucker. In a mean-field approach, we assume that the 
density of solute particles can be obtained from the time-dependent analytical solution of the free-space diffusion 
equation with a point source (the chucker), integrated over the previous trajectory of the chucker. 
This approach assumes that the thermal fluctuations of the chucker are slower with respect to the solute diffusion, and further neglects thermal fluctuations in the solute density.
%This approach 
%assumes a separation of timescales between the thermal fluctuations of  the chucker and the solute: we include the 
%contribution to the chucker dynamics of the mean solute concentration field (produced during a given dynamical 
%trajectory of the chucker), but  neglect contributions of thermal fluctuations in the solute density. 
A similar approach has recently been used to simulate a microorganism performing chemotaxis by Sengupta {\em{et al}} \cite{Lowen09}.

In our Langevin dynamics simulations, the force $\bf{f}$ on the ``point chucker'' depends on the concentration 
profile $c({\bf{r}})$ of the solute according to~\cite{Gol05,Gol07,And1,And2}:
\begin{equation}
\label{eq:velgrad}
{\bf{f}} = - A \nabla c({\bf{r}})
\end{equation}
where the prefactor $A$ is determined by the physical process by which an osmotic pressure gradient is converted 
into colloidal motion. This force produces a drift velocity  ${\bf{v}}$ which is related to the force ${\bf{f}}$ 
by ${\bf{f}}=k_BT {\bf{v}}/D_1 = 6\pi \eta R_1 {\bf{v}}$. By tuning the prefactor $A$, we can tune the drift velocity ${\bf{v}}$. 
If the solvent-chucker interactions were to be properly taken into account, linearised theory suggests that the drift velocity would be 
given by  Golestanian's relation ${\bf{v_s}}=-k_BT \lambda^2 \nabla c / \eta$ \cite{Golestanian09}, where (as discussed earlier) $\lambda$ 
is related to the lengthscale of solvent-colloid interactions. By combining this relation with Eq.(\ref{eq:velgrad}), we can obtain an ``effective''
 value of $\lambda$ for our Langevin dynamics simulations:  $\lambda=\sqrt{A/(6\pi R_1 k_BT)}$. A full treatment, including solvent, with 
this value of $\lambda$, would give the same drift velocity as our Langevin dynamics simulations with prefactor $A$. By varying the 
prefactor $A$, we can vary the effective $\lambda$ value. By choosing $\lambda$ values corresponding to  realistic solvent-colloid 
interaction lengths, we should obtain results closer to those that would be expected for a full treatment including solvent. In contrast, 
in our Monte Carlo simulations, the value of $\lambda$ is fixed. Here,  the net force on the chucker produced by the imbalance of solute 
collisions across the colloid surface \cite{Brady08} corresponds to a value $A=(4/3)\pi R_1^3 k_BT$, or alternatively an effective $\lambda = \sqrt{2}R_1/3$.

% We note that $A=(4/3)\pi R_1^3 k_BT$ would be also expected for the case of a  totally solvent-permeable colloid, with hydrodynamic effects. 

The concentration profile $c({\bf{r}})$ is not represented explicitly in our simulations but is calculated analytically 
from the time history of the position of the point chucker:
\begin{equation}\label{eq:one}
c({\bf{r}}) = k_{\rm c} \int_{0}^{t} dt' G({\bf{r-r'}},t-t'),
\end{equation}
where 
${\bf{r}} \equiv {\bf{r(t)}}$, ${\bf{r'}} \equiv {\bf{r(t')}}$ and $G({\bf{r-r'}},t-t')$ is the 
Green's function for the diffusion equation in free space with a sink at infinity (valid for comparison to our MC simulations when the MC boundary 
position $L$ is large):
\begin{eqnarray}\label{eq:green}
G({\bf{r-r'}},t-t') = \frac{1}{8(\pi D_2 (t-t'))^{3/2}} \exp{\left[-\frac{(x-x')^2+(y-y')^2+(z-z')^2}{4D_2(t-t')}\right]}
\end{eqnarray}
where $D_2$ is the diffusion constant for the solute. We now define $F({\bf{r-r'}},t-t') = \nabla G({\bf{r-r'}},t-t')$. Since in our Langevin 
dynamics algorithm the particle moves in discrete steps, we can replace the integral in Eq.(\ref{eq:one}) by a sum of $F({\bf{r-r'}},t-t')$ over the previous trajectory of the particle, 
giving us a history-dependent expression for the force on the particle:
\begin{equation}\label{eq:sum}
{\bf{f}}(t) = -A k_c \sum_{i=1}^{M}  \Delta t F\left({\bf{r}}(t)-{\bf{r}}(t-i\Delta t),i\Delta t\right)
\end{equation}
where $\Delta t$ is the timestep. The sum in Eq.(\ref{eq:sum}) is carried out only over the previous $M$ steps, 
where $M\Delta t$ should be longer than  the timescale over which the solute diffuses away. The equation of motion for the point chucker is then:
\begin{equation}\label{eq:mot}
{\bf{r}}(t+\Delta t) = {\bf{r}}(t) + \frac{D_1}{k_BT}\Delta t{\bf{f}}(t) + {\bf{\xi}}
\end{equation}
%CHECK THIS! 
where $D_1$ is the diffusion constant of the chucker and  ${\bf{\xi}}$ represents Gaussian white 
noise with zero mean and variance $2D_1\Delta t$. Care is needed when selecting the parameters $D_1$, $D_2$ and $k_c$ 
for comparability of the results of the Langevin and MC simulations, because of the different time units of the two algorithms; however, 
dimensionless combinations of parameters can be defined which are equivalent between the two simulation schemes.

In our MC simulations, we expect that at high chucking rates, the solute particles will tend to form a cage around 
the chucker, restricting its diffusion. This effect is not included in the Langevin dynamics simulations where we 
consider a point chucker and a continuous solute field. The Langevin Dynamics simulations are therefore only valid for low chucking 
rates, where the solute concentration is small.

%However, we can incorporate into our Langevin dynamics 
%simulations a dependence of the viscosity $\eta$ on the concentration of solute: this will tend to slow the chucker 
%down when the local solute concentration is high. To achieve this, we assume that our solute particles are polymer 
%coils (inspired by bacterial exopolysacchide) and use the non-linear relation for polymer  suspensions \cite{einstein_viscosity}:
%\begin{equation}\label{eq:viscosity}
%\eta(\phi) = \eta_0\frac{1+0.5\phi}{(1-\phi)^2},
%\end{equation}
%where $\eta_0$ is the solvent viscosity and $\phi$ is the volume fraction 
%of solute particles, which can be calculated from the Green's function (\ref{eq:green}), assuming a particular solute radius $R_2$:
%\begin{equation}
%\phi(t) = \frac{4\pi R_2^3}{3}\sum_{i=1}^{M} G\left({\bf{r}}(t)-{\bf{r}}(t-i\Delta t),i\Delta t\right) 
%\end{equation}
%Eq.(\ref{eq:viscosity}) describes the increase in viscosity as the solute packing fraction increases. Combining Eq.(\ref{eq:viscosity}) with the 
%Stokes-Einstein relation \cite{einstein}, we obtain an expression for the diffusion coefficient 
%$D_1$ of the chucker:
%\begin{equation}\label{eq:se}
%D_1=\frac{k_BT}{6 \pi R_1 \eta(\phi)}.
%\end{equation}
%This expression is evaluated at each timestep and used in the equation of motion (\ref{eq:mot}). We note that 
%the physics behind this assumed that the increase of viscosity is quite different from that underlying the solute caging in our 
%MC simulations. There is therefore no reason to suppose that our Langevin and MC simulation results will coincide for high chucking rates.

\section{Results}\label{sec:results}
In Sections~\ref{sec:conc} and~\ref{sec:diff}, we first discuss the  behaviour of a chucker in free space, with no external applied forces. 
We then determine the response of the chucker to a dragging force in Section~\ref{sec:drag}, and 
finally we discuss the behaviour of a chucker which sediments against an absorbing or hard-planar surface in Section~\ref{sec:wall}. 

\subsection{Solute concentration profiles}\label{sec:conc}

We begin by investigating the concentration profile of solute particles around the chucker in our MC simulations when $R_1/R_2=10$, 
for a fixed chucker. Since the solute particles do not interact with each other, we expect the steady-state 
solute concentration profile to correspond to the solution of the diffusion equation (\ref{eq:one}), using the Green's 
function appropriate for a sink at distance $L$ from the chucker. 
This solution (for the solute concentration $c$) 
is $c(r) = \frac{k_{\rm c}}{4 \pi D_2}\left( \frac{1}{r} - \frac{1}{L}\right)$, so that the solute volume fraction 
$\phi$ is given by  $\phi(r) = \frac{k_{\rm c}R_2^3}{3 D_2} 
\left( \frac{1}{r} - \frac{1}{L}\right)$. We note that this is not a true volume fraction, since solute particles do not interact 
and can overlap in our model (so that $\phi$ can be greater than one); but it does give some indication of the density of the solutes, for comparison to real systems.

\begin{figure}[h!]
\begin{center}
\includegraphics[width=9cm,clip]{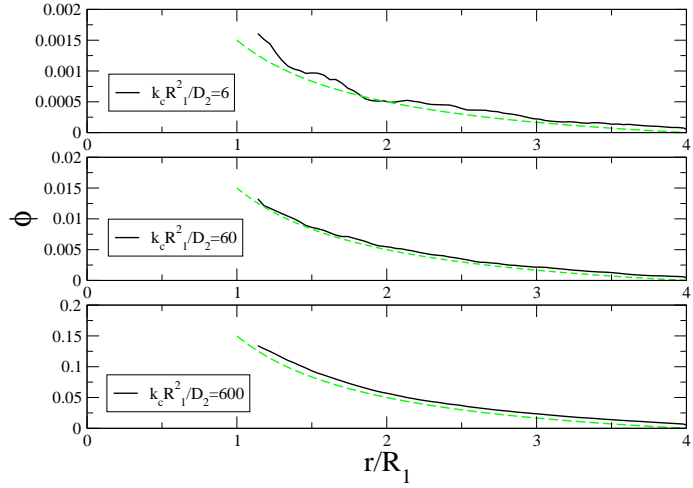}
\caption{{Comparison between analytical expression (green dashed-line) and simulated data (solid
black line) for the ``volume fraction'' $\phi \equiv (4/3)\pi R_2^3 c$ (where $c$ is concentration) of solute particles surrounding a non-motile chucker ($R_1/R_2=10$) with different values 
of $k_{\rm c}$ (from top to bottom, $k_{\rm c}=$  0.00001, 0.0001, 0.001 [MC cycles]$^{-1}$). The corresponding values of 
the dimensionless ratio $k_{\rm c}R_1^2/D_2$ are given in the legend, for comparison 
%with the Figure~\ref{fig:diffusion} 
with the analytical expression. Note that since the solute particles can overlap, $\phi$ is not a true packing fraction.\label{fig:theorysim}}}
\end{center}
\end{figure}

Figure~\ref{fig:theorysim} shows a comparison between the radial profiles 
for $\phi$ obtained from our simulations with the analytical expression, for $L=4 R_1$. The simulation data agrees 
very well with the analytical expression; the small  discrepancy close to the boundary ($r/R_1=4$) occurs because 
the analytical expression is computed for a  spherical boundary whereas the simulation has a cubic boundary.

\begin{figure}[h!]
\begin{center}
\includegraphics[width=9cm,clip]{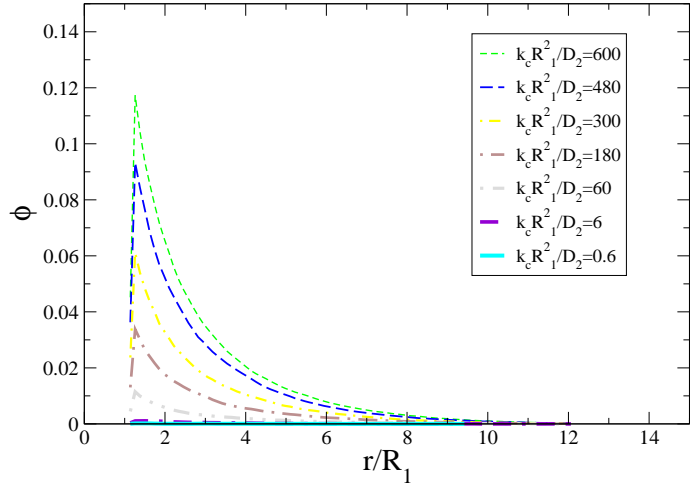}
\caption{``Volume fraction'' $\phi \equiv (4/3)\pi R_2^3 c$ (where $c$ is concentration) of solute particles surrounding the chucker ($R_1/R_2=10$) as a function of the 
distance from the centre of the chucker (expressed in units of $R_1$). The corresponding values of 
the dimensionless ratio $k_{\rm c}R_1^2/D_2$ are given in the legend (note that  the solute particles can overlap).
\label{fig:packing-fraction}}
\end{center}
\end{figure}

We are also interested in how closely the solute particles are packed around 
the chucker: if the volume fraction close to the chucker surface 
is too high then the the Asakura-Oosawa approximation fails to accurately describe a real solute (e.g. polymer coils), since it neglects solute-solute interactions. 
Figure~\ref{fig:packing-fraction} shows the volume fraction $\phi$ of solute particles in spherical shells centred 
on the chucker in our MC simulations, computed for a freely diffusing chucker with $R_1/R_2=10$, over 
the full range of chucking rates used in our simulations ($ 0 <  k_c R_2^2 /D_1\le 600 $). The maximum volume fraction obtained close to the 
chucker is about 0.12, and for most chucking rates it is much lower. This suggests that our neglect of solute-solute 
interactions is justified over the range of parameters used in this work.

\subsection{Chucker diffusion}\label{sec:diff}
We now discuss the diffusive behaviour of an isolated chucker, using MC simulations with two different ratios 
of the chucker and solute radii: $R_1/R_2=10$ and $R_1/R_2=100$, for various chucking rates $k_c$. In our simulations, 
the solute particles are produced isotropically so there is no net directional motion. Rather, we observe diffusive 
behaviour, in which the mean square displacement $\langle r(t)^2 \rangle$ is linear in time. This allows us to compute 
an effective diffusion constant   $D_{\rm{eff}}$ by linear fitting of the long-time behaviour of $\langle r(t)^2 \rangle$.

Figure~\ref{fig:diffusion} shows that in our simulations the effective diffusion constant $D_{{\rm{eff}}}$ of the chucker varies 
non-monotonically with the  chucking rate $k_c$. Diffusive motion of the chucker is enhanced 
as the chucking rate increases, for low chucking rates.  At high chucking rates, diffusive motion is inhibited due to 
crowding by the solute particles, so that $D_{{\rm{eff}}}$ decreases with $k_c$.  We plot the dimensionless quantity 
$D_{{\rm{eff}}}/D_1$, where  $D_1$ is the diffusion constant of the chucker for zero chucking rate, versus $k_{\rm c}R^2_2/D_1$. The 
latter turns out to be the most relevant dimensionless parameter for comparison to theory: see Ref~\cite{Golestanian09} and Eq.(\ref{eq:gol1}). 
$R_2^2/D_1$ is the typical time for the chucker to diffuse the radius of a solute particle; this ratio therefore measures the relative magnitudes of this time and the inverse of the chucking rate.
%which is  the ratio between the time the chucker takes to 
%diffuse its own radius ($R^2_1/D_1$) and the typical time  $1/k_{\rm c}$ to generate a new solute particle. 

We believe that the mechanism behind the increase in $D_{{\rm{eff}}}$ with chucking rate, for small $k_c$, is as follows. A stationary chucker 
would be surrounded by a symmetric,  concentration field of solute which decreases with distance from the chucker. A small 
random displacement of the chucker, however, positions it asymmetrically in the solute concentration field.
%, so that the solute 
%concentration in front of the chucker is less than that behind it.
 Osmotic effects will then tend to cause the chucker to move 
further in the same direction, leading to a  non-thermal osmotic enhancement of the effective diffusion constant of the chucker.

Comparing the results in Figure~\ref{fig:diffusion} for the two different ratios of chucker 
and solute radius, we find that the small $k_c$ regime, where $D_{\rm{eff}}/D_1$ increases with $k_c R_2^2 / D_1$,  appears not to 
depend strongly on the size of the solute -- although the inset, plotted on a linear scale, suggests that the two plots do not actually collapse convincingly onto one curve. On the other
hand, our data suggests that the regime, for large $k_c$, in which diffusion decays with chucking rate, does
depend strongly on the size of the crowder: crowding effects kick in at smaller $k_c R_2^2 / D_1$ for the smaller solute particles. 
This suggests that the dependence of the crowding effect on solute radius is less steep than $\sim R_2^2$. 

Golestanian's recent theoretical work \cite{Golestanian09}, which uses linearised hydrodynamics, predicts 
that $D_{\rm{eff}}/D_1$ should be given by:
\begin{equation}\label{eq:gol1}
\frac{D_{\rm{eff}}}{D_1}=1+\left[1.17810 \left(\frac{\lambda}{R_1}\right)^4  \left(\frac{3}{\pi}\right)\right]\left(\frac{k_c R_2^2}{D_1}\right)
\end{equation}
 where  we have substituted the 
Stokes-Einstein relation ($D_1=\frac{k_BT}{6 \pi R_1 \eta}$) and the relation $\mu=k_BT\lambda^2/\eta$ into the expression given in Eq.(7) of Ref \cite{Golestanian09}. 
In our MC simulations, the imbalance of solute collisions across the chucker surface is expected to 
produce a net force which corresponds to a prefactor $A=(4/3)\pi R_1^3 k_BT$ in Eq.(\ref{eq:velgrad}). This will 
produce a drift velocity ${\bf{v}} = {\bf{f}}/(6 \pi \eta R_1)$. Linearized theory, which accounts for solvent-colloid 
surface interactions, predicts a colloid velocity ${\bf{v_s}}=-k_BT \lambda^2 \nabla c / \eta$ \cite{Golestanian09}. Matching 
these two relations, we find that our MC simulations have an effective $\lambda$ value $\lambda=\sqrt{2}R_1/3$. Substituting this 
into Eq.(\ref{eq:gol1}), we obtain a prediction for $D_{\rm{eff}}$ in our MC simulations.
%In the absence of hydrodynamics, as discussed above, we might expect that $\lambda=\sqrt{2}R_1/3$, corresponding to a 
%prefactor $A=(4/3)\pi R_1^3 k_BT$ in Eq.(\ref{eq:velgrad}).

\begin{figure}[h!]
\begin{center}
\includegraphics[width=9cm,clip]{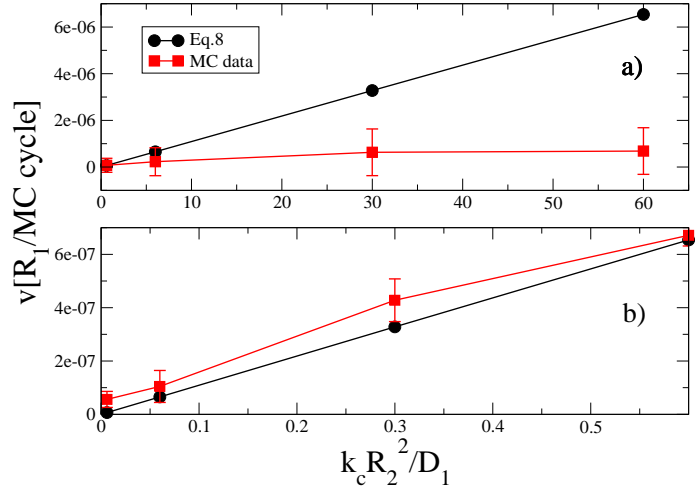}
\caption{Sedimentation velocity for a passive colloid in a
concentration gradient of solute, generated by a planar source and absorbed at a planar sink. The dimensionless combination $k_c R_2^2/D_1$ 
measures the solute concentration gradient (note the colloid does not chuck in these simulations; solute is introduced into the simulation
 box at the source plane with rate $k_c$). The black circles indicate analytical results from Eq.(\ref{velgrad}), whereas the
red squares are simulation data. (a): the case $R_1/R_2=10$ (b): the case $R_1/R_2=100$.\label{fig:grad10-100}}
\end{center}
\end{figure}

 The inset to Figure~\ref{fig:diffusion} shows a comparison between this prediction and our MC data, for small chucking rates. For
 $R_1/R_2=100$, we obtain an excellent fit, whereas for the larger solute particles,  $R_1/R_2=10$, the fit is much less convincing. 
We will revisit the effect of solute size in Figure \ref{fig:grad10-100}. 
We note that Golestanian \cite{Golestanian09} also predicts a regime in which the mean square displacement $\langle r^2 \rangle$ is 
proportional to $t^{3/2}$, for timescales between the inertial and diffusive regimes, for an isotropic chucker. We do not observe 
this regime in our simulations: for ratios of $R_1/R_2 \ge 5$ the mean square displacement is linear in time. 
%Interestingly, this breaks down when $R_1$ approaches $R_2$, where we find evidence of sub-diffusive dynamics.

\begin{figure}[h!]
\begin{center}
\includegraphics[width=9cm,clip=true]{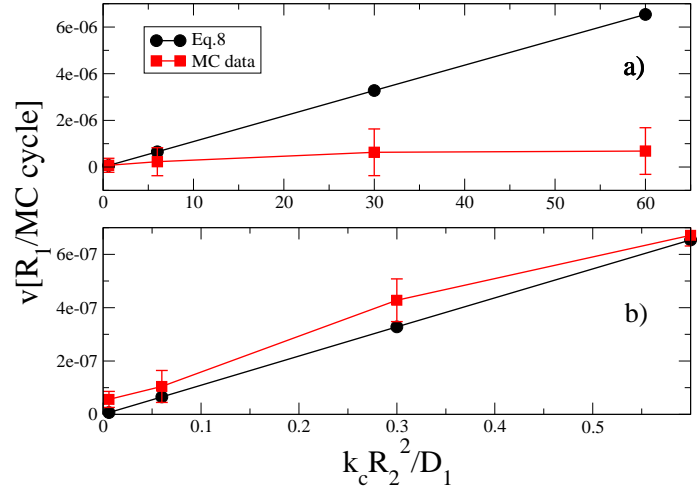}
\caption{Snapshot of a passive colloid in a concentration gradient of Asakura-Oosawa solute particles: 
solute particles are generated at the top planar source, and removed at the bottom planar 
sink (the positions of these planes are indicated by the blue circles). \label{fig:gradsnap}}
\end{center}
\end{figure}

To further investigate the relationship between chucker diffusion and solute concentration 
gradient, we performed additional MC simulations for a {\em passive} colloid,
of radius $R_1$, in a constant density gradient of Asakura-Oosawa solute particles
(Fig.\ref{fig:gradsnap}). In order to maintain a constant density in steady state, 
we introduced a planar source of solute particles (of radius $R_2$) at the top of the
simulation box, and a planar absorbing sink at the bottom. Solute particles are randomly seeded on the source plane at a constant rate $k_c$.

%Periodic boundary conditions are applied in the plane of the source and sink.

The analysis in Ref.~\cite{Gol05} can be generalised to
yield the following prediction for the steady state velocity of a colloidal
particle in a constant concentration gradient (neglecting hydrodynamic effects):
\begin{equation}
\label{velgrad}
v = \frac{4 \pi R_2 R_1^2 k_c}{3 L^2}
\end{equation}
where the top and bottom 
planes have a surface area of $L^2$, and we have again  
used the expectation that $\lambda=(\sqrt{2}/3)R_1$ in our
simulations. 

Figure \ref{fig:grad10-100} shows a comparison between the analytical prediction (\ref{velgrad}) and our simulation results, 
for $R_1/R_2=10$ (panel a) and $R_1/R_2=100$ (panel b). As in the inset to Figure \ref{fig:diffusion}, the agreement with the theoretical prediction is much better for the smaller
solute radius. This observation is not unexpected, given that the smaller solute radius is closer to the
continuum limit.
What is perhaps surprising is the extent of the deviation from the theory when
$R_1/R_2=10$. For this solute radius, our simulation results deviate from the analytical prediction by about an order of magnitude.

An important drawback of our MC simulation scheme is that it does not account for the details of the solvent interactions with the chucker surface.
  As discussed in Section~\ref{sec:intro}, 
these effects may drastically reduce the magnitude of solute-induced effects on colloid mobility. In our MC simulations, an 
imbalance of solute concentration across the chucker surface leads to fewer overlaps (``collisions'') on one side of the 
chucker than the other, and hence net motion down the concentration gradient. We expect this to 
correspond to Eq.(\ref{eq:velgrad}) with a prefactor of  $A=(4/3) \pi R_1^3 k_BT$ (simply summing collisions across the colloid surface). 
We would expect hydrodynamic effects at the chucker surface (if included) to decrease this prefactor: for a solvent-impermeable chucker with no 
slip boundary conditions, linearised hydrodynamic theory predicts that the prefactor decreases  by a factor $\sim \lambda^2/R_1^2$ 
where  $\lambda$ [the lengthscale for chucker-solute interactions] is  $\sim R_2$ \cite{And1,And2,Golestanian09,Gol05}. 
If, on the other hand, the  chucker is partially or totally permeable to solvent, the decrease in the prefactor due to these effects 
should be less severe \cite{Gol05,sucrose}.

\begin{figure}[h!]
\begin{center}
\includegraphics[width=9cm,clip]{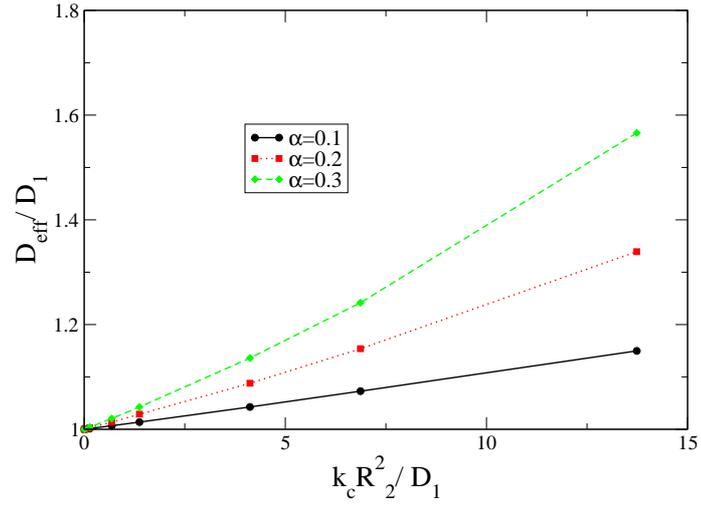}
\caption{$D_{\rm{eff}}$ computed with Langevin dynamics simulations.
The Langevin dynamics simulations are
performed with various prefactors $A=\alpha k_B T R_1^3$, where
$k_B T=1$, and the values of $\alpha$ are given in the
legend.\label{fig:comparison}}
%\label{MC_LD_1}
\end{center}
\end{figure}

As discussed in Section \ref{sec:lang}, we can model the effects of solvent-chucker surface interactions approximately using 
Langevin dynamics simulations in which we vary the magnitude of the prefactor $A$: this corresponds to varying the hydrodynamic lengthscale $\lambda$. Figure~\ref{fig:comparison} shows the 
effective diffusion constant $D_{\rm eff}$, 
computed as a function of $k_c$ using Langevin dynamics simulations with different prefactors, for $R_1/R_2=10$. The same qualitative 
behaviour is observed as in our MC simulations (compare to Figure \ref{fig:diffusion}): again, $D_{\rm eff}/D_1$ increases with 
$k_c$ for small chucking rates. For large chucking rates, the effective diffusion constant decreases in the MC simulations due to 
crowding effects which are not included in the Langevin Dynamics simulations
%shows a peak at intermediate $k_c$. Although the magnitude of this peak decreases as the prefactor decreases, the peak persists 
%even for rather small prefactors (at least 20 times smaller than the naive kinetic value $(4/3)\pi R_1^3 k_BT$), suggesting that the 
%qualitative behaviour predicted by our MC simulations is robust 
%to hydrodynamic effects. 

\begin{figure}[h!]
\begin{center}
\includegraphics[width=9cm,clip=]{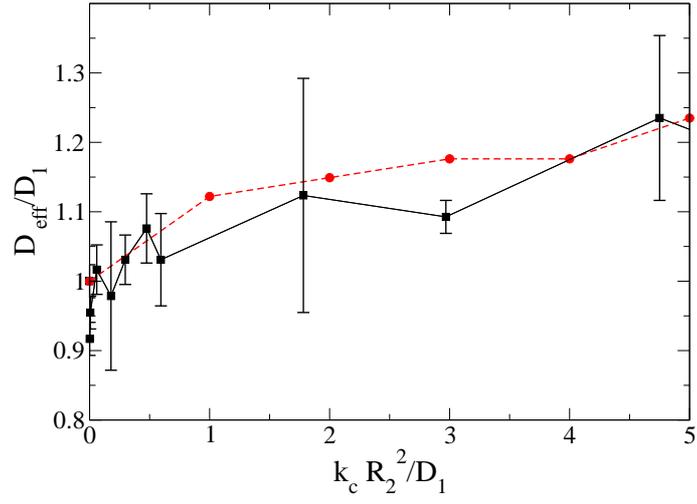}
\caption{$D_{\rm{eff}}/D_1$ versus the dimensionless ratio $k_{\rm c}R^2_2/D_1$ for a 
chucker with $R_1/R_2=100$: the black continous line represents the MC data and the red dashed line the Langevin 
results with $A=(4/3)\pi R_1^3 k_BT$. The plot shows data only for small values of $k_{\rm c}R_2^2/D_1$. 
\label{fig:MCLD_100}}
\end{center}
\end{figure}

For small $k_c R_2^2/D_1$, we can make a quantitative comparison between the MC and Langevin simulation results, as shown in Figure \ref{fig:MCLD_100}.  
Here, the Langevin simulations were run with the ``naive kinetic'' prefactor  $A=(4/3)\pi R_1^3 k_BT$. 
Excellent agreement is obtained between the two types of simulation. This reinforces our earlier conclusions 
that the MC simulations for small $R_2$ behave as would be predicted by hydrodynamic theory with a hydrodynamic 
lengthscale $\lambda = (\sqrt{2}/3)R_1$. 
%As discussed earlier, quantitative comparison between the MC and Langevin simulations is not possible for large $k_c$, because the physical 
%origin of the decrease in $D_{\rm eff}$ is different in the two simulation schemes. 

\subsection{A dragged chucker}\label{sec:drag} 

For passive systems, such as colloidal particles, the fluctuation-dissipation theorem (FDT) states that the mobility coefficient $\Gamma$ is 
fundamentally related to the diffusion constant $D$ by $\Gamma=D/k_BT$. For active systems, there is no reason why the FDT should hold. 
Deviations from the FDT not only serve to demonstrate that the system is out of equilibrium, but can also yield insight into the underlying physics. 

To investigate the relationship between mobility and diffusion for the colloidal chucker, we carried out  MC simulations in which  a 
chucker (with $R_1/R_2=10$)  is dragged in a fixed direction by a constant force $F$ 
(for example, using optical tweezers). We measure the velocity $v$ of the chucker in the direction of the force. The mobility coefficient 
$\Gamma$ is defined by $v=\Gamma F$; this can be compared with the value of $D_{\rm eff}$ obtained in Figure \ref{fig:diffusion}. In this 
work, we use values of $F$ which approximately correspond to the magnitude of the gravitational 
force on a bacterium.

\begin{figure}[h!]
\begin{center}
\includegraphics[width=9cm,clip=]{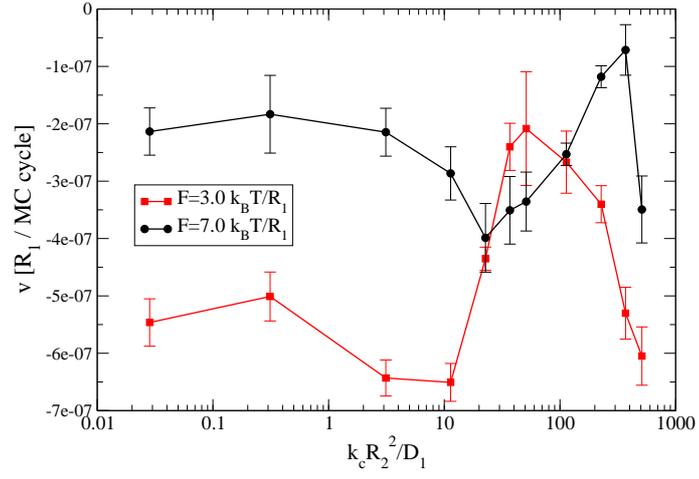}
\caption{Asymptotic velocity $v$ versus the dimensionless ratio $k_{\rm c}R^2_2/D_1$ for a 
chucker under an external force of $F=3.0$ $k_BT/R_1$ or $F=7.0$ $k_BT/R_1$ ($R_1/R_2=10$).
\label{fig:drag1}}
\end{center}
\end{figure}

Figure~\ref{fig:drag1} shows the steady state chucker velocity $v$, as a function of the chucking rate, expressed as  the 
dimensionless ratio $k_{\rm c}R^2_2/D_1$, for two different values of the pulling force $F$. This velocity depends  strongly and 
nonmonotonically on the chucking rate, in a markedly different way to the effective diffusion constant of Figure \ref{fig:diffusion}.

Combining the results of Figures \ref{fig:diffusion} and \ref{fig:drag1}, we obtain Figure \ref{fig:drag2}, which describes how the ratio  
$\Gamma k_B T/D_{\rm eff}$ depends on  $k_{\rm c}R^2_2/D_1$. Deviation of $\Gamma k_B T/D_{\rm eff }$ 
from unity (green dashed line) corresponds to violation of the FDT. We observe firstly that deviations from FDT in this system are strong 
(up to a factor of 3) and nonmonotonic. Depending on the chucking rate and the pulling force, mobility may dominate diffusion, or vice versa. 
As might be expected, the deviations from FDT are stronger for the larger pulling force.  While it is not surprising that the FDT 
does not hold in this nonequilibrium system, what is perhaps unexpected is that the relationship between the mobility and diffusion 
constants has such a complex dependence on both the chucking rate and the applied force for this rather simple model.

The trends observed in Figure \ref{fig:drag2} can be tentatively related to changes in the solute configuration around the chucker, for different chucking rates and pulling forces. 
Naively, one might imagine that for small chucking rates, where the density of solutes is low, 
the pulling force might tend to displace the chucker relative to its symmetric concentration 
gradient of solute, so that more solute would be behind it than in front, 
resulting in an osmotic enhancement of the mobility. This is what one would expect from an advection-diffusion equation with a source moving at a finite
Peclet number (see e.g. Ref.~\cite{goldstein}). However, for large chucking rates, the density of solutes near the chucker is higher, and one might 
imagine that the pulling force would lead to an increased collisions of the chucker with nearby solute particles, decreasing the mobility.

To test this hypothesis, we plot in Figure \ref{fig:cone} the ``volume fraction'', as defined by $\phi \equiv (4/3)\pi R_2^3 c$,  of solute particles 
within the conical volume supported by the angle $\theta = \pi/9$, in front of and behind the chucker 
(the cone is extended all the way to the edge of the box). We first note that there is a difference in the volume fraction of solute in front 
of and behind the chucker, for large enough chucking rate. These effects are more pronounced for the larger pulling force. For   $F=7.0 k_BT/R_1$, 
the solute concentration is larger behind than in front of the chucker, for chucking rates in the range $50 < k_c R_2^2/D_1 < 200$. 
We would expect this to cause an enhancement of mobility over diffusion; however, this range of $k_c$ values is beyond the peak in $\Gamma k_B T/D_{\rm eff}$ 
(as shown by the green dashed line). For higher chucking rates, we would expect to see a larger volume fraction in front of the chucker, 
if the solute particles are inhibiting chucker motility. There is some indication that this happens for the highest chucking rates
(not shown in Figure \ref{fig:cone}).
It is possible that this crowding effect plays a more important role for the smaller value of the force $F=3.0 k_BT/R_1$, 
where the pulling force is not strong enough to overcome the crowding; for the stronger pulling force $F=7.0 k_BT/R_1$, we speculate that the pulling 
force is strong enough to overcome this effect.  Further work will be needed to fully understand the data in Figure \ref{fig:drag2}; it is clear, 
however, that the nontrivial relationship between mobility and diffusion in this system is caused by a complex interplay between crowding and osmotic effects.

\subsection{A sedimenting chucker}\label{sec:wall}

We now consider the case of a chucker which experiences a constant force $F$ perpendicular to a planar surface which either 
absorbs the solute particles or behaves as a hard wall. This models the gravity-driven sedimentation of a chucker onto a surface, in the 
case where gravitational effects on the solute particles are negligible.
Although this model is highly simplified, the sedimentation of exopolysaccharide-producing bacteria onto surfaces is important
 in the formation of biofilms \cite{sutherland2001}. Here, the hard surface case might mimic the initial stages of biofilm 
formation, when bacteria sediment onto an uncolonised surface, while the absorbing surface  might correspond to a situation 
where bacteria sediment onto an already formed biofilm  which tends to absorb exopolysaccharide.  

We determine the steady-state probability distribution function for the chucker position relative to the surface in our MC simulations. 
This would correspond to an experimental sedimentation density profile, in the case of low chucker density where interactions between 
chuckers can be ignored.  
In all our simulations, we take  $R_1/R_2=10$, $F=7.0 k_BT/R_1$ and initialise the simulations 
with the centre of the colloid at a distance $z=1.6 R_1$ from the surface; we then begin to observe the system once a steady state has been 
reached  (i.e. after $9 \times 10^{5}$ Monte Carlo cycles). This choice of $F$ approximately 
corresponds to the gravitational force experienced by a typical bacterium.

\begin{figure}[h!]
\begin{center}
\includegraphics[width=9cm,clip=true]{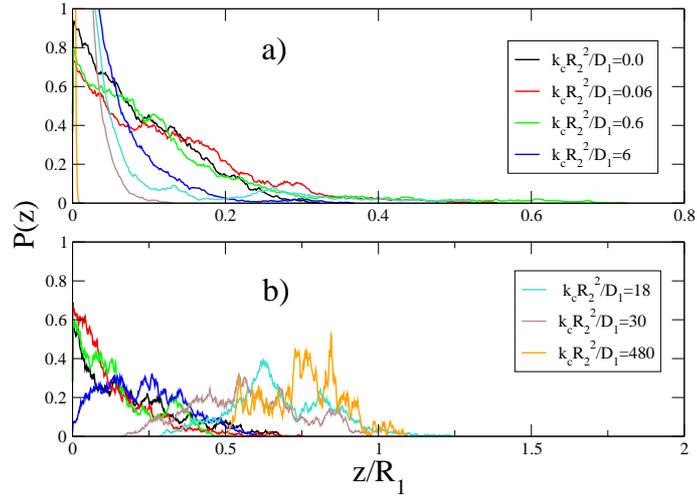}
\caption{Steady-state probability distribution for the $z$-coordinate of the centre of the sedimenting chucker, 
for $R_1/R_2=10$ and various chucking rates (indicated in the legend).\label{fig:wall}}
\end{center}
\end{figure}

Figure~\ref{fig:wall} shows the steady-state probability distribution  $P(z)$ for the position of the centre of the chucker 
relative to the surface, which is at $z=0$. In Figure~\ref{fig:wall}(a), the wall acts as a sink for solute particles. 
This results in a buildup of the chucker probability density close to the surface: the higher the chucking rate, the more 
tightly the chucker tends to approach the surface. This effect can be understood in terms of an effective attractive interaction 
between the chucker and the surface, in which local depletion of solute near the surface causes an osmotic imbalance which squeezes 
the chucker against the surface. Figure~\ref{fig:wall}(b) shows the corresponding probability density profiles when the planar surface 
instead acts as a hard wall. Here we find two regimes, depending on the chucking rate. For small chucking rates ($k_{\rm c}
R_2^2/D_1\le 1$), 
the probability density decreases monotonically away from the surface, with 
a steepness that decreases as $k_c$ increases (for $k_c=0$ we obtain a Boltzmann distribution). However, as $k_c$ increases further, the 
profile develops a minimum close to the surface and a peak at some distance from the surface. This peak recedes from the surface with 
increasing $k_c$. For the hard surface, there is a solute-mediated osmotic repulsion between the chucker and the surface. The 
crossover between the two types of probability density profile occurs when this repulsion becomes strong enough to overcome the 
gravitational force pulling the chucker towards the surface. In the high $k_c$ regime, the peak in $P(z)$ occurs where the 
gravitational and osmotic repulsion forces balance. We note that for the hard wall, the solute density profile, and hence the steady-state 
chucker location, depends on the position of the absorbing box boundary, since a linear concentration gradient of solute arises between the 
colloid and the box boundary. However, the qualitative features of our results are independent of the boundary position.

\section{Discussion}\label{sec:discussion}

In this work, we have used computer simulations to study  the physics of an 
active colloid (a ``chucker'') of radius $R_1$, which produces solute particles of radius $R_2$ isotropically on its  surface, at a rate  $k_{\rm c}$. 
Most of our simulations have used a Monte Carlo scheme in which the chucker is modelled as a hard sphere and the 
solute particles as Asakura-Oosawa ({\em{i.e.}} mutually non-interacting) spheres. Our MC simulations were 
complemented by an overdamped  Langevin  dynamics approach in which the chucker is represented as a point particle 
which experiences a force proportional to the gradient of a concentration field of solute, this field being computed 
analytically using the Green's function for free diffusion. The Langevin dynamics approach has the advantage that we 
can tune the prefactor that links the osmotic force to the solute concentration gradient: this provides a crude way 
to model hydrodynamic effects close to the chucker surface.

%alterations in the hydrodynamic properties of the chucker (e.g. its solvent permeability or surface slip velocity). 

Our MC simulations of an isolated chucker in free space show diffusive behaviour, from which we extract an 
effective diffusion constant $D_{\rm{eff}}$. This varies nonmonotonically with the chucking rate. For small 
values of $k_c$, solute chucking enhances diffusion. We interpret this as being due to the chucker being ``pushed along'' 
by a self-generated solute concentration gradient; the same behaviour is recovered in our Langevin dynamics simulations. 
An increase in $D_{\rm{eff}}$ as a function of $k_c$ was recently
 predicted analytically by Golestanian \cite{Golestanian09}. In our MC simulations, when the chucking rate becomes large, the chucker 
tends to become caged by newly generated solute particles, hampering its diffusion. The result is therefore a peak 
in $D_{\rm{eff}}$ at intermediate chucking rate. In our simulations, the solute particles do not interact. For more realistic, 
interacting solutes, we might expect crowding to kick in at lower chucking rates, shifting the peak in $D_{\rm{eff}}$ towards smaller $k_c$.

% Langevin dynamics simulations produce qualitatively 
%the same behaviour, although in this case the decrease in $D_{\rm{eff}}$ at large $k_c$ is not due to 
%caging but rather to an assumed increase in solvent viscosity at high solute concentration. 

A quantitative comparison between our MC simulation results and Golestanian's theoretical prediction reveals that the simulations are in 
agreement with theory for a hydrodynamic length $\lambda = (\sqrt{2}/3)R_1$, for large values of the ratio
between the sizes of the chucker and of the solute particles, $R_1/R_2=100$; agreement is less convincing for $R_1/R_2=10$. This is the value 
of $\lambda$ we would predict by matching the drift velocity predicted in a  ``naive'' osmotic view, in which the colloid moves because of 
an imbalance in the number of solute collisions across its surface, with the colloid velocity, expressed as a function of $\lambda$, predicted by hydrodynamic theory. 
When we simulate a passive colloid in an externally imposed concentration gradient of solute particles, we also find good agreement with the 
hydrodynamic theory prediction with  $\lambda = (\sqrt{2}/3)R_1$ for  $R_1/R_2=100$, but for $R_1/R_2=10$, the  diffusiophoretic mobility is
 smaller than might naively be predicted from osmotic pressure arguments. We speculate that this poor agreement for large solutes may be due to 
local fluctuations of the solute concentration field, neglected in the theory. We can also make a quantitative comparison, for small chucking rates, 
 between our MC and Langevin simulation results. For  $R_1/R_2=100$, the two simulation schemes produce effective diffusion constants in good 
agreement, for a choice of Langevin prefactor of $A=(4/3) \pi R_1^3 k_BT$, corresponding to $\lambda = (\sqrt{2}/3)R_1$.

\begin{figure}[h!]
\centerline{
\includegraphics[width=9cm,clip]{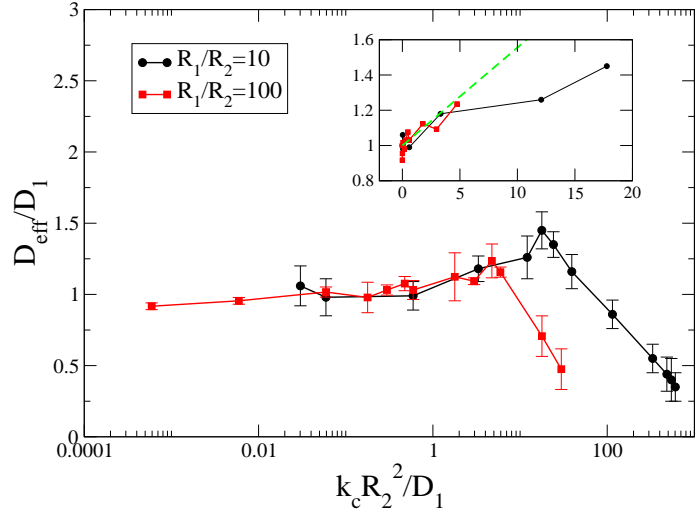}}
\caption{Effective diffusion coefficient $D_{\rm{eff}}$ of a chucker
(normalised by the diffusion constant in the absence of chucking $D_1$)
as a function of the dimensionless ratio $k_{\rm c}R^2_2/D_1$, for
$R_1/R_2=10$ (black circles) and
$R_1/R_2=100$ (red squares). Note the logarithmic scale on the horizontal axis. The inset (on a linear scale) shows a comparison between the
same data for small $k_c$ and the theoretical prediction,
Eq.(\ref{eq:gol1}),
with $\lambda=\sqrt{2} R_1/3$ (green dashed line).\label{fig:diffusion}}
\end{figure}

\begin{figure}[h!]
\begin{center}
\includegraphics[width=9cm,clip=]{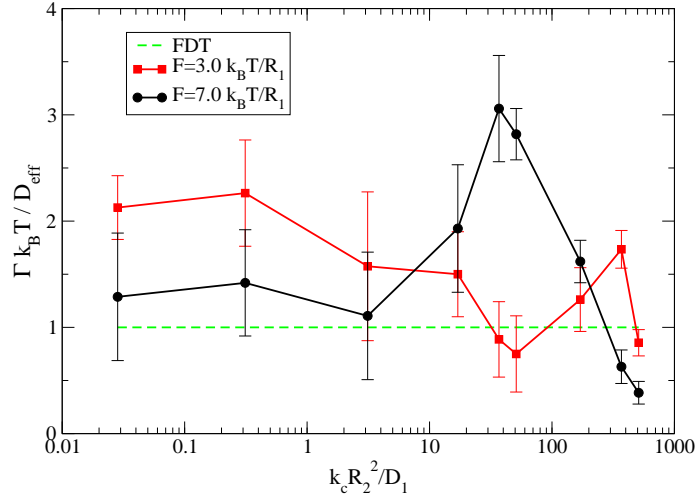}
\caption{$\frac{\Gamma k_B T }{D_{\rm eff}}$ 
versus the dimensionless ratio $k_{\rm c}R^2_2/D_1$, for a chucker with $R_1/R_2=10$. $D_{\rm{eff}}$ is measured from the 
mean square displacement of a freely diffusing chucker (Figure \ref{fig:diffusion}) whereas $\Gamma$ is obtained by dragging the chucker with an external 
force of $F=3.0 k_BT/R_1$ or $F=7.0 k_BT/R_1$ (Figure \ref{fig:drag1}). 
%UNITS? FIX AXIS LABELS
The dashed line indicates the FDT prediction for a passive system.\label{fig:drag2}}
\end{center}
\end{figure}

\begin{figure}[h!]
\begin{center}
\includegraphics[width=9cm,clip=]{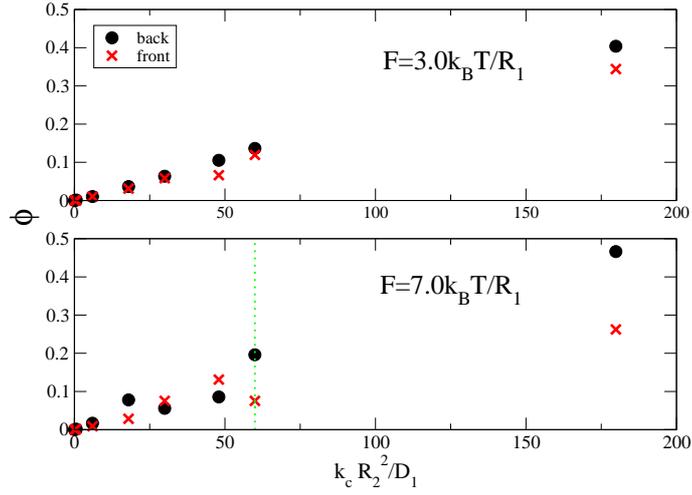}
\caption{``Volume fraction'' $\phi \equiv (4/3)\pi R_2^3 c$ (where $c$ is concentration) of solute within a cone of angle 
$\theta = \pi/9$ at the front (red crosses)  
and at the back (black dots) of the pulled chucker, as a function of the dimensionless ratio $k_{\rm c}R^2_2/D_1$, 
for a chucker with $R_1/R_2=10$  pulled by an external force of $F=3.0 k_BT/R_1$ or $F=7.0 k_BT/R_1$. 
The green dotted line represents the position of the peak of $\Gamma k_B T/D_{\rm eff}$ in Figure \ref{fig:drag2}.
For $F=3.0 k_BT/R_1$, the  peak of $\Gamma k_B T/D_{\rm eff}$ is around 360 and not shown.
 Note also that the solute particles can overlap.
\label{fig:cone}}
\end{center}
\end{figure}

One of the initial motivation for this study was to investigate possible effects of exopolysaccharide 
production on bacterial diffusivity. The bacterium {\em X. campestris} produces xanthan polymer at a rate as high as $10^4$ polymer 
coils per cell per second (as estimated from the data given in  Refs.~\cite{xanthomonas1,xanthomonas2}). Assuming a ratio $R_1/R_2=10-100$, corresponding to a polymer gyration radius of 10-100 nm,
hence to a value of $k_{\rm c}R_2^2/D_1\sim 1-100$, and referring to
Fig.~\ref{fig:diffusion}, we find that the effective diffusion constant of the chucker should be significantly different from the infinite dilution value
at this chucking rate.  On the other hand, the bacterium {\em{Sinorhizobium meliloti}}, a less prolific producer of exopolysaccharide, produces  
less than 1 polymer per bacterium per second~\cite{gary}, leading to a value of $k_{\rm c}R_2^2/D_1$ of $10^{-4}-10^{-2}$ (assuming the same dimensions for the polymer), so that for 
this bacterium, we expect the effect of polymer production on diffusion to be essentially perturbative. It would be interesting 
to compare the diffusion constants of mutants of {\em X. campestris} which are altered in their exopolysaccharide production rates: 
our simulations suggest a change in $D_{\rm{eff}}$ of up to about 50\% if exopolysaccharide production were to be eliminated. 
One would need to be careful in such an experiment to avoid the buildup of polymer in the bacterial suspension, 
leading to crowding and a subsequent decrease in diffusion.

The chucker is an example of a nonequilibrium, ``active'' system, in which statistical mechanical relations such as 
the Fluctuation-Dissipation theorem (FDT) may not hold. Comparing the effective diffusion constant  $D_1$ 
to a mobility coefficient computed by dragging a chucker with a constant force in our MC simulations, we find that
 indeed this system violates the FDT strongly, in a highly nontrivial way. The complex relationship between mobility and diffusion in this 
system is due to an interplay between crowding and osmotic effects which depends sensitively on the applied force and chucking rate.

Finally, we have studied the behaviour of a chucker which sediments onto a planar surface which either absorbs the solute particles or 
behaves as a hard wall. When the surface is absorbing, a solute-mediated effective attraction (which 
increases with increasing $k_c$) pushes the chucker towards the surface. However, when we consider a hard  surface, 
a solute-mediated effective repulsion exists. For small $k_c$ this simply results 
in a decrease in the gradient of the chucker probability density profile near the surface. For large $k_c$, the repulsive 
interaction dominates the gravitational force near the surface, resulting in  a  peak in the probability density for the chucker position at some distance from the surface.

This work was motivated by recent experimental, theoretical and simulation work on  colloidal ``swimmers'' 
which catalyse a chemical reaction, or secrete some product, anisotropically across their 
surface~\cite{osmoGol07,Sen1,Sen2,Golestanian09,Gol05,Brady08,Kapral,popescu}, by recent theoretical 
work on isotropic chuckers \cite{Golestanian09} and by the biological example of exopolysaccharide-secreting 
bacteria \cite{sutherland1982,sutherland1994}. However, the physical mechanisms explored here also have 
wider relevance in soft matter physics. As an example, Jiang {\em{et al}} \cite{Wada09} recently reported a system where 
migration of colloidal particles was mediated by  inhomogeneities in polymer concentration, which were 
in turn induced by a temperature gradient. Dynamic depletion interactions between colloidal particles, 
in which one of the components of a phase separating binary fluid wets the colloid surface (creating a 
local depletion near the colloid), predicted in simulations by Araki and Tanaka \cite{araki}, are also 
closely related to the phenomena studied here. 

We hope that the results reported here will make a useful contribution to 
ongoing discussions concerning the effects of osmotic gradients on colloidal dynamics, as 
well as linking these discussions to the physics of biological systems (particularly exopolysaccharide-producing bacteria) and other nonequilibrium soft matter systems.

{\em Acknowledgements:} We thank M. E. Cates, O. Croze, G. Dorken, G. Ferguson, D. Frenkel, 
W. C. K. Poon, E. Sanz, P. Visco and L. Wilson for discussions and 
advice, and the referee for comments that have significantly improved the paper. This work was supported by EPSRC under grant EP/E030173/1. 
CV acknowledges support from a Marie Curie Fellowship. 
RJA was funded by the Royal Society of Edinburgh and by the Royal Society.
This work has made use of the resources provided by the Edinburgh Compute and 
Data Facility (ECDF). The ECDF is partially 
supported by the eDIKT initiative.

\bibliography{chucker}

\newpage 

\newpage 

\newpage

%\begin{figure}[h!]
%\begin{center}
%\includegraphics[width=9cm,clip]{Figures/plot-confronto-phi-con-theory.6.eps}
%\caption{Comparison between analytical expression (green dashed-line) and simulated data (solid black line) for the ``volume fraction'' 
%{\red{$\phi \equiv (4/3)\pi R_2^3 c$ (where $c$ is concentration)}} of solute particles surrounding a non-motile chucker ($R_1/R_2=10$) with different values 
%of $k_{\rm c}$ (from top to bottom, $k_{\rm c}=$  0.0001, 0.001 [MC cycles]$^{-1}$). The corresponding values of 
%the dimensionless ratio $k_{\rm c}R_1^2/D_2$ are given in the legend, for comparison 
%%%with the Figure~\ref{fig:diffusion} 
%with the analytical expression. {\red{Note that since the solute particles can overlap, $\phi$ is not a true packing fraction.}}\label{fig:theorysim}}
%\end{center}
%\end{figure}

\newpage 

%\begin{figure}[h!]
%\centerline{
%\includegraphics[width=9cm,clip]{Figures/packing-fraction.21.5.eps}}
%\caption{``Volume fraction'' {\red{$\phi \equiv (4/3)\pi R_2^3 c$ (where $c$ is concentration)}} of solute particles surrounding the chucker ($R_1/R_2=10$) as a function of the 
%distance from the centre of the chucker (expressed in units of $R_1$). The corresponding values of 
%the dimensionless ratio $k_{\rm c}R_1^2/D_2$ are given in the legend {\red{(note that  the solute particles can overlap).}}
%\label{fig:packing-fraction}}
%%The corresponding values of the dimensionless ratio $k_{\rm c}R_1^2/D_1$ are:  30000, 18000, 6000, 4800, 
%%3000, 1800, 600, 60 and 6. \label{fig:packing-fraction}}
%%for comparison with Figure~\ref{fig:diffusion}.\label{fig:packing-fraction}}
%\end{figure}

\newpage

%\begin{figure}[h!]
%\begin{center}
%\includegraphics[width=9cm,clip]{Figures/diff_vs_kon.final.6.eps}
%\caption{Effective diffusion coefficient $D_{\rm{eff}}$ of a chucker
%(normalised by the diffusion constant in the absence of chucking $D_1$)
%as a function of the dimensionless ratio $k_{\rm c}R^2_2/D_1$, for
%$R_1/R_2=10$ (black circles) and
%$R_1/R_2=100$ (red squares). Note the logarithmic scale on the horizontal axis. The inset (on a linear scale) shows a comparison between the
%same data for small $k_c$ and the theoretical prediction,
%Eq.(\ref{eq:gol1}),
%with $\lambda=\sqrt{2} R_1/3$ (green dashed line).\label{fig:diffusion}}
%\end{center}
%\end{figure}

\newpage 

\newpage 

\newpage 

\newpage 

\newpage 

\newpage 

\newpage 

\newpage 

\end{document}